\documentclass[11pt]{article}

\usepackage[noend]{algpseudocode}
\usepackage{amsfonts}
\usepackage{amsmath}
\usepackage{amssymb}
\usepackage{amsthm}
\usepackage{authblk}
\usepackage{booktabs}
\usepackage{caption}
\usepackage{subcaption}
\usepackage{color}
\usepackage[a4paper,margin=1in]{geometry}
\usepackage{graphicx}
\usepackage[]{hyperref}
\hypersetup{
    colorlinks = true,
    allcolors = black
}
\usepackage[utf8]{inputenc}
\usepackage[numbers,sort&compress]{natbib}
\usepackage{pgfplots}
	\pgfplotsset{compat=1.12}
\usepackage{tikz}
	\usetikzlibrary{shapes,plotmarks}

\newcommand{\zero}{\emph{Type~0}~}
\newcommand{\one}{\emph{Type~1}~}
\newcommand{\two}{\emph{Type~2}~}
\newcommand{\three}{\emph{Type~3}~}
\newcommand{\four}{\emph{Type~4}~}
\newcommand{\five}{\emph{Type~5}~}
\newcommand{\six}{\emph{Type~6}~}

\definecolor{myRed}{RGB}{224, 24, 24}

\begin{document}

\title{
	Adversaries monitoring Tor traffic crossing their jurisdictional border 
	and reconstructing Tor circuits
}
\author[]{Herman Galteland\thanks{This work is funded by Nasjonal 
sikkerhetsmyndighet (NSM), \url{www.nsm.stat.no}.} }
\author[]{Kristian Gjøsteen}
\affil[]{
	Department of Mathematical Sciences, \\ Norwegian University of 
	Science and Technology, NTNU \\
	\texttt{\small
		$\{$herman.galteland, kristian.gjosteen$\}$@ntnu.no
	}
}
\maketitle

\begin{abstract}
We model and analyze passive adversaries that monitor Tor traffic crossing the border of a jurisdiction the adversary is controlling. We show by simulations that a single jurisdiction is able to connect incoming and outgoing traffic of their border, tracking the traffic, and a coalition of jurisdictions is able to reconstruct parts of the Tor network, revealing user-website connections. We use two algorithms to estimate the capabilities of the adversaries, the first simulates a Tor network and the second analyzes data from the simulation and reconstructs the network.
\end{abstract}


\section{Introduction}

The Onion Router (Tor) protocol~\cite{TOR} is a well-established onion routing system that tries to provide a low-latency communications channel while also defending against network-level adversaries trying to reveal who is talking to whom. It is well understood how the Tor network behaves when an adversary compromises a fraction of the onion routers, and in particular that if the entire network is monitored little or no security is left.

In this paper we analyze the power of (coalitions of) less powerful adversaries who do not monitor onion router traffic directly, but instead partition the network into jurisdictions and monitor traffic crossing from one partition into another.

These kinds of adversaries are interesting because they are real, in particular of the form of programs to monitor \emph{traffic crossing borders}. In 2008 the Swedish parliament passed a bill allowing the Swedish National Defence Radio Establishment (\emph{Försvarets radioanstalt}) to monitor both wireless and cable signals passing the Swedish border~\cite{Sweden}. Denmark~\cite{Denmark}, France~\cite{France}, and the United Kingdom~\cite{UnitedKingdom} all have similar laws on how to gather and store digital information. 

In 2016, the Norwegian government appointed a group of experts to investigate whether or not the Norwegian Intelligence Service should be allowed access to communication crossing the Norwegian border, similar to the Swedish National Defence Radio Establishment. The investigating report concluded that the Norwegian Intelligence Service should be allowed to monitor the Norwegian border~\cite{DGF}, however, this has not yet been put into effect. It seems likely that other countries either have or plan to have similar programs.

\subsection{Related work}

Formal analysis of the Tor protocol comes in two variants. The first use an abstract model of the protocol and gives security bounds based on a worst case adversary~\cite{ANOA, EDG09, FJS07, FJS12, GH13, HM08}. The second includes a detailed description of the protocols in their analysis when proving the security bound~\cite{JWJSS13, WTSB13, BKMM14}. 

Adversaries that observe both ends of a Tor circuit can connect the user with the website it is communicating with~\cite{MD05, OS06}. The literature has considered adversaries controlling: an \emph{autonomous system} (AS)~\cite{FD04, SEVLRCM15}; an \emph{Internet exchange point} (IXP)~\cite{MZ07}; and several ASes and IXPs~\cite{JWJSS13, NSGZS16}. It has been shown that ASes can observe both ends of Tor circuits~\cite{WTSB13}. Tor path selection algorithms has been proposed to avoid detection from ASes~\cite{AYM14, ES09}.

An adversary with access to timing, packet size, and directionality of packets sent over an encrypted HTTP tunnel can reveal the identity of the server and user by \emph{traffic analysis} attacks~\cite{D02, BLJL06, H02, LL06, SSWRPQ02, CLSFG08}. Countermeasures to traffic analysis attacks includes \emph{padding messages}~\cite{RFC5246} and \emph{morphing Tor traffic} to mimic traffic associated to other servers~\cite{CLSFG08}. Note that hiding the packet length is insufficient~\cite{DCRS12}.

\emph{Tor network simulators}~\cite{ShadowTor, torps} makes it possible to analyze the effectiveness of adversaries versus the Tor protocol.

A \emph{stepping stone} is an intermediate node used by an attacker to conceal his identity. Algorithms used to detect stepping stones analyses streams of traffic to confirm or reject the existence of intermediate nodes between the analyzed traffic streams~\cite{WRW02, BSV04}.

\subsection{Our contribution}

In this paper we model and discuss a specific adversary versus the Tor protocol. The \emph{jurisdictional adversary} is similar to an adversary controlling AS(es) or IXP(s), however, the ASes and IXPs are typically located inside a jurisdiction whereas we consider a passive adversary that only monitors traffic crossing the border of a jurisdiction. Further, an adversary controlling an AS or an IXP would see all traffic inside their network whereas an adversary monitoring jurisdictional borders would not.

We simulate a Tor network, which includes the adversaries monitoring and storing traffic crossing their border. A chosen coalition of jurisdictions is trying to reconstruct the simulated Tor network by analyzing the stored data using traffic analysis. We do not morph the Tor traffic since the 
adversaries are only interested in the existence of traffic and not what it 
looks like.

Algorithms used to detect stepping stones analyzes streams of traffic to find intermediate nodes between the streams. Similarly, our reconstruction algorithm attempts to connect stream of traffic between known onion routers to recreate circuits. The techniques used to detect stepping stones could be used to detect onion routers. The difference between the stepping stone literature and our work is the adversary we are modeling and analyzing, where we assume that the location of all onion routers is already known and we want to connect Tor traffic to reconstruct Tor circuits.

\subsection{Overview}

The model for our overlay network of Tor is in Section~\ref{sec: Modeling jurisdictional adversaries}, this model is used to classify the types of traffic, and connections, the jurisdictional adversaries can observe, and create. The simulation algorithm is described in Section~\ref{sec: Simulation} and the reconstruction algorithm in Section~\ref{sec: Reconstruction}. In Section~\ref{sec: Reconstruction results} we present the reconstruction test results. We conclude with a possible countermeasure against the adversaries and summarize the adversaries in Section~\ref{sec: Disscussions}. We include the parameters used for the reconstruction results in Appendix~\ref{app: Simulation and reconstruction parameters}.

\section{Background}
\label{sec: Background}

\subsection{Tor}
\label{sec: Tor}

The Onion Router protocol is an anonymous communication protocol~\cite{TOR}. The Tor protocol uses intermediate nodes called \emph{onion routers} to achieve anonymity. A user establishes a \emph{circuit} of onion routers in the Tor network to communicate with a server, where each onion router only knows the identity of its neighboring nodes. The first onion router of a circuit is a \emph{guard node} $G$ and when a user creates a new circuit he picks the guard node from a small set of onion routers, the default Tor configuration is three guard nodes. The last onion router communicates with the server on behalf of the user and is called an \emph{exit node} $E$. The Tor protocol does not ensure encryption between the exit and the server and the a malicious actor could abuse the information. Only a few onion routers get marked as an exit and it is believed that the majority of the exit nodes are honest. The intermediate node, between the guard and exit, is a \emph{relay node} $R$. We let \emph{circuit node} refer to any of the nodes in a circuit. For simplicity we assume that all circuits consist of one user, three onion routers, and one server, which is the default Tor configuration. Restricting circuits to contain only three onion routers is not essential for our reconstruction algorithm.

The user establishes a secret key with each onion router in the circuit and encrypts data in layers when sending it to the server, where each onion router removes one layer of encryption before relaying the data to the next node. When the server sends data back to the user each onion router encrypts the data and adds a layer. Note that we are only interested in the flow of information and will not include any encryption in our simulation.

\section{Modeling jurisdictional adversaries}
\label{sec: Modeling jurisdictional adversaries}

We describe the overlay network of the Tor network and use this model to determine what types of traffic a jurisdiction could observe.

\subsection{Overlay network}

The \emph{overlay network} of the Tor network describes how information is sent between circuit nodes. A node in the overlay network represents a jurisdiction and an edge represents a communication connection between two jurisdictions. A jurisdiction contains a number of circuit nodes, where we assume the jurisdictions know which node is located inside its border. Information is sent in the overlay network when circuit nodes communicate. If two communicating circuit nodes are located in the same jurisdiction then no information is sent, and if the two circuit nodes are located in two different jurisdictions a \emph{network path} in the overlay network is chosen. This path consists shows how the traffic is sent between jurisdictions, from the first jurisdiction containing the sender circuit node to the last jurisdiction containing the receiver circuit node, and determines which jurisdiction is able to observe the traffic data sent between the two circuit nodes.

Note that we make a simplification. Traffic between two circuit nodes inside a jurisdiction could very well cross the jurisdiction's borders in the physical network. In fact, since routing is dynamic, it could cross borders one day and not cross borders the next. Hence, the adversaries get less information in our model than in the real world.

\subsection{Observable traffic}
\label{sec: Observable traffic}

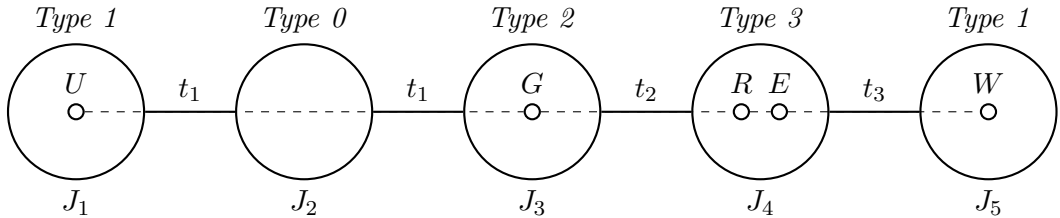
\begin{figure}[t]
\centering
\begin{tikzpicture}
	\tikzstyle{leaf}=[circle,draw,inner sep=2pt,thick]
	\tikzstyle{jur}=[circle,draw,inner sep=18pt,thick]

	\node[leaf,label={90:{$U$}}] (U) at (-6,0) {};
	\node[leaf,label={90:{$G$}}] (G) at (0,0) {};
	\node[leaf,label={90:{$R$}}] (R) at (2.75,0) {};
	\node[leaf,label={90:{$E$}}] (E) at (3.25,0) {};
	\node[leaf,label={90:{$W$}}] (S) at (6,0) {};
	
	\node[jur,label={270:{$J_1$}},label={90:{\one}}] (j1) at (-6,0) {};
	\node[jur,label={270:{$J_2$}},label={90:{\zero}}] (j2) at (-3,0) {};
	\node[jur,label={270:{$J_3$}},label={90:{\two}}] (j3) at (0,0) {};
	\node[jur,label={270:{$J_4$}},label={90:{\three}}] (j4) at (3,0) {};
	\node[jur,label={270:{$J_5$}},label={90:{\one}}] (j5) at (6,0) {};
	
	\draw[dashed] (U) -- (G) -- (R) -- (E) -- (S); 
	\draw[thick] (j1) --node[draw=none, above] {$t_1$} (j2) --node[draw=none, above] {$t_1$} (j3) --node[draw=none, above] {$t_2$} (j4) --node[draw=none, above] {$t_3$} (j5); 
\end{tikzpicture}
\caption{An example of the overlay network with a Tor circuit. $U$ denotes a user, $G$ a guard node, $R$ a relay node, $E$ an exit node, $W$ a website, $J_1$, \dots, $J_5$ denotes five distinct jurisdictions, and $t_1, t_2, t_3$ denotes timestamps for packets traveling between two circuit nodes. $J_1$ and $J_5$ observe \emph{Type 1} traffic, an endpoint. $J_2$ observe \emph{Type 0}, traffic passing through. $J_3$ observe \emph{Type 2} traffic, where an incoming and an outgoing packet share a common node $G$ and the timestamp difference $|t_1 - t_2|$ is close to an expected value. $J_4$ observe \emph{Type 3} traffic, where the observed incoming and outgoing packets do not share a node but the timestamp difference $|t_2 - t_3|$ is close to an expected value. The solid line shows the network paths and the dashed line shows the Tor circuit}
\label{fig: Type 0-3}
\end{figure}

As a Tor user communicates with a website they both generate traffic data. The user sends data packets to the first node of the circuit, which forwards it to the next node in the circuit and so forth until the website receives the user's packets, similar for the website. Packet data sent between two circuit nodes are transferred over a network path and all jurisdictions in the path observe the packets' metadata information. We assume a jurisdiction learns the sender, receiver, and direction of the packet and has the timestamp for when it observed the packet. A packet is observes when it crosses the border of a jurisdiction. A packet can be incoming, entering the jurisdiction, outgoing, leaving the jurisdiction, or passing through a jurisdiction.

The jurisdictional adversaries want to reconstruct the Tor circuits to reveal the sender and user, breaking the \emph{relationship anonymity}~\cite{PH09} of the Tor protocol. Using the observed packet data a jurisdiction can combine incoming and outgoing packets using traffic analysis. When an onion router receives a packet it will either encrypt or decrypt it, to add or remove an onion layer. This cryptographic computation takes time and there will be a timestamp difference between the observed incoming and outgoing packet. If the timestamp difference is close to an expected value then the two observed incoming and outgoing packets are most likely part of the same circuit, which means they can be connected. The time it takes to send a packet over a network cable is negligible compared to the time it takes for a circuit node to do its cryptographic computations and we assume that sending packets over a cable takes no time at all. We classify the types of connections a jurisdiction can create from its observed packets into four categories, see Figure~\ref{fig: Type 0-3} for a visual description;
\begin{description}
\itemsep=-2.5pt
\item[\zero] A single packet passing through the jurisdiction, where the sender and receiver node of the packet is not located inside the jurisdiction's borders.
\item[\one] A single packet ending in the jurisdiction, where either the sender or the receiver node of the packet is located inside the jurisdiction and is an endpoint of the circuit (a user or a website).
\item[\two] One incoming and one outgoing packet share a common node inside the jurisdiction and the packets' timestamp difference is close to an expected value.
\item[\three] One incoming and one outgoing packet that do not share a common node, but their timestamp difference is close to an expected value.
\end{description}
Packets that can be connected are combined and stored as \emph{partial circuits}, each partial circuit contains a path of circuit nodes representing a partial Tor circuit and timestamps. We say a partial circuit has length $n$ if its path consists of $n$ circuit nodes. The timestamps are collected from the packet(s) it is created from, and all timestamp of any packet which would make the same partial circuit. (Many similar packets each with one timestamp makes one partial circuit with many timestamps.) The timestamps may be ordered into different sets to show the direction and flow of traffic over the partial circuit. 

\subsection{Reconstructable traffic}
\label{sec: Reconstructable traffic}

\begin{figure}[t]
\centering
\begin{subfigure}[t]{0.31\textwidth}
\begin{tikzpicture}
	\tikzstyle{leaf}=[circle,draw,inner sep=2pt,thick]
	\tikzstyle{jur}=[circle,draw,inner sep=18pt,thick]
	
	\tikzstyle{user}=[leaf,label={90:{$U$}}]
	\tikzstyle{guard}=[leaf,label={90:{$G$}}]
	\tikzstyle{relay}=[leaf,label={90:{$R$}}]
	\tikzstyle{exit}=[leaf,label={90:{$E$}}]
	\tikzstyle{website}=[leaf,label={90:{$W$}}]

	\node[draw=none] at (2, 1.5) {\four};
	\node[user] (4U) at (0,0) {};
	\node[guard] (4G) at (1,0) {};
	\node[relay] (4R1) at (2,0) {};
	\node[exit] (4E1) at (3,0) {};
	
	\node[relay] (4R2) at (2,-1) {};
	\node[exit] (4E2) at (3,-1) {};
	\node[website] (4W) at (4,-1) {};
	
	\draw[thick] (4U) -- (4G) -- (4R1) --node[draw=none, above] {$t_1$} (4E1);
	\draw[thick] (4R2) --node[draw=none, above] {$t_2$} (4E2) -- (4W);
\end{tikzpicture}
\caption{Two partial circuits that share two nodes, and there are enough timestamps that are equal $t_1~=~t_2$.}
\end{subfigure}
\begin{subfigure}[t]{0.31\textwidth}
\begin{tikzpicture}
	\tikzstyle{leaf}=[circle,draw,inner sep=2pt,thick]
	\tikzstyle{jur}=[circle,draw,inner sep=18pt,thick]
	
	\tikzstyle{user}=[leaf,label={90:{$U$}}]
	\tikzstyle{guard}=[leaf,label={90:{$G$}}]
	\tikzstyle{relay}=[leaf,label={90:{$R$}}]
	\tikzstyle{exit}=[leaf,label={90:{$E$}}]
	\tikzstyle{website}=[leaf,label={90:{$W$}}]
		
	\node[draw=none] at (7, 1.5) {\five};
	\node[user] (5U) at (5,0) {};
	\node[guard] (5G) at (6,0) {};
	\node[relay] (5R1) at (7,0) {};
	
	\node[relay] (5R2) at (7,-1) {};
	\node[exit] (5E) at (8,-1) {};
	\node[website] (5W) at (9,-1) {};
	
	\draw[thick] (5U) -- (5G) --node[draw=none, above] {$t_1$} (5R1);
	\draw[thick] (5R2) --node[draw=none, above] {$t_2$} (5E) -- (5W);
\end{tikzpicture}
\caption{Two partial circuits that share one node, and there are enough timestamps differences $|t_1 - t_2|$ that are close to an expected value.}
\end{subfigure}
\begin{subfigure}[t]{0.31\textwidth}
\begin{tikzpicture}
	\tikzstyle{leaf}=[circle,draw,inner sep=2pt,thick]
	\tikzstyle{jur}=[circle,draw,inner sep=18pt,thick]
	
	\tikzstyle{user}=[leaf,label={90:{$U$}}]
	\tikzstyle{guard}=[leaf,label={90:{$G$}}]
	\tikzstyle{relay}=[leaf,label={90:{$R$}}]
	\tikzstyle{exit}=[leaf,label={90:{$E$}}]
	\tikzstyle{website}=[leaf,label={90:{$W$}}]
		
	\node[draw=none] at (12, 1.5) {\six};
	\node[user] (6U) at (10,0) {};
	\node[guard] (6G) at (11,0) {};
	\node[relay] (6R) at (12,0) {};
	
	\node[exit] (6E) at (13,0) {};
	\node[website] (6W) at (14,0) {};
	
	\node[draw=none] at (12, -1) {};
	
	\draw[thick] (6U) -- (6G) --node[draw=none, above] {$t_1$} (6R);
	\draw[thick] (6E) --node[draw=none, above] {$t_2$} (6W);
\end{tikzpicture}
\caption{Two partial circuits that do not share any nodes, however, there are enough timestamps differences $|t_1 - t_2|$ that are close to an expected value.}
\end{subfigure}
\caption{Examples of partial circuits and how they can be connected. $U$ denotes a user, $G$ a guard node, $R$ a relay node, $E$ an exit node, $W$ a website, and $t_1, t_2$ denotes timestamps for packets traveling between two circuit nodes.}
\label{fig: Type 4-6}
\end{figure}
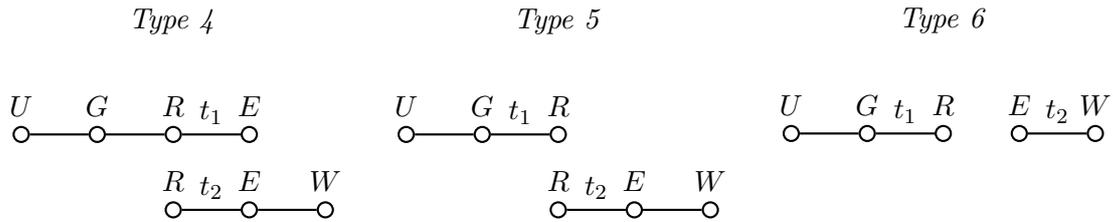

It is very unlikely that a single jurisdiction is able to reveal the sender and receiver of any circuit. A coalition, however, can combine their analyzed partial circuits and potentially create full Tor circuits, breaking the relationship anonymity. The colluding jurisdictions share their partial circuits with each other and try to combine them. We want to track the packets traveling from one jurisdiction to the next and look for partial circuits with paths that overlap, such that we can make a longer path by combining them. We classify the types of connections a coalition can create from its partial circuit, see Figure~\ref{fig: Type 4-6} for a visual description;
\begin{description}
\itemsep=-2.5pt
\item[\four] Two partial circuits with paths that share two common nodes, where the two nodes are located at the end of the first and at the beginning of the second, and there are enough timestamp pairs, one from each partial circuit, that are identical.
\item[\five] Two partial circuits with paths that share a common node, where that node is located at the end of the first and at the beginning of the second, and there are enough timestamp pairs, one from each partial circuit, that have a timestamp difference that is close to an expected value.
\item[\six] Two partial circuits with paths that do not share any common nodes, but there are enough timestamp pairs, one from each partial circuit, that have a timestamp difference that is close to an expected value.
\end{description}
Two partial circuits which combines into a \four connection should have identical timestamps. They share two nodes and the timestamps sent between these two nodes are should be present in both partial circuits.

Note that \five and~\emph{6} connections are similar to \two and~\emph{3} connections, the only difference is that partial circuits are being connected instead of single packets and in the reconstruction algorithm we reuse many of the methods for connecting these types of traffic.

\section{Simulation algorithm}
\label{sec: Simulation}

We describe our algorithm simulating the Tor network. The flow of information is sent in our overlay network, detailed in Section~\ref{sec: Modeling jurisdictional adversaries}. In our simulator we assume that the adversaries are able to recognize Tor traffic and only generate Tor traffic, since onion routers usually only send Tor traffic to each other and all onion routers are known. Further, we assume that the adversaries have analyzed the distribution of timing patterns of Tor traffic. They will use this knowledge to statistically connect traffic entering and exiting their jurisdiction. 

We initiate a network of jurisdictions, place onion routers, and define the user and website distributions. When a user communicates with a website, creates a new circuit, or destroys circuits we generate data. All traffic generated is sent between circuit nodes in the form of packets. Any data crossing the border of a jurisdiction is observed and stored, either as incoming or outgoing traffic. This data will be used in the reconstruction algorithm, detailed in Section~\ref{sec: Reconstruction}.

We try to simulate the real world as best as we can, where information about the Tor network is gathered from the Tor data analysis website ``Tor metrics''~\cite{TorMetrics}.

\subsection{Initializing an overlay network}
Each initialized jurisdiction represents a real world country. Two jurisdictions are connected by an edge if they share a border or if they connected to the same underwater internet cable~\cite{SubmarineCables}.

Guard, Relay, and Exit nodes are placed in jurisdictions, where the location of each node is gathered from Tor metrics~\cite{TorMetrics}.

Users and websites are not placed in a jurisdiction at initialization, they are chosen and created during the simulation. The distribution of Tor users is gathered from Tor metrics~\cite{TorMetrics} and a user can communicate with a website from any jurisdiction, hence, the websites are uniformly distributed.

\subsection{Generate packets}
The simulation runs for $n$ iterations and each iteration generates data for a user that communicates with a website.

We pick either an existing ready user or create a new. If there is a ready user, we create a new circuit if its existing has been up for more that 10 minutes. If there is no ready user we make a new user with a fresh circuit. Creating circuits generate traffic data.

When a user and website communicate they sends data over the circuit, where each node in the circuits sends packets to each other. A packet includes a timestamp, the sender and receiver node, the circuit ID of the Tor circuit, and packet length. It has the form
\[
	\texttt{Timestamp Sender > Receiver (Circuit ID) Length}.
\]
Note that the circuit ID is only used to verify the output of the reconstruction algorithm. 

\subsection{Time and timestamps}
A global \texttt{TIME} parameter is used to maintain the order of the user's activity, it is increased by a positive value between each iteration. The global parameter stays fixed during an iteration while the active user's time continues. As a packet is forwarded in the circuit the nodes does cryptographic operations on it, although we do not do the actual computations we add an \emph{onion router delay} to the users time. Similarly, we add a \emph{sender delay} between packets sent from the user and website. 

A lognormal distribution is used to sample these delays. Each onion router use its own distribution to compute its delays, and it is sampled from a family of lognormal distributions. We do not know which distribution each onion router is using. We only know the family of distribution, which is chosen such that the circuit round-trip latency of the simulated network is close to the real Tor network's latency~\cite{TorMetrics}.

When the user has finished its activity in the current iteration it is getting ready for its next by waiting, as if reading the website it is communicating with. An activity delay is added to the user's time, which is uniform. Whenever the \texttt{TIME} parameter is larger than the user's time activity delay it can be chosen as a ready user.

\subsection{Write observed traffic data}
Jurisdictions in the network path between two circuit nodes observes and stores the traffic. Each jurisdiction writes data to file as either incoming or outgoing traffic, this data will be used in the reconstruction algorithm. 

\subsection{Runtime}

The runtime of the simulation algorithm is mainly dominated by writing and storing all data generated, the number of iterations $n$ determines how much data is generated. For each iteration we generate traffic data for one Tor user, as it communicates with a website. Using the simulation parameters in Table~\ref{tab: Simulation parameters}, one iteration will on average generate 14,000 packets. However, more that one jurisdiction observes each packet and on average 60,000 packets are stored each iteration.

In wall-clock runtime for the simulation algorithm for: $n = 1000$ is roughly 5 minutes, $n = 10000$ is 4 hours, and $n = 100000$ is 11 hours.

\section{Reconstruction algorithm} 
\label{sec: Reconstruction}

We describe our reconstruction algorithm, where a coalition of jurisdictions wants to reconstruct Tor circuits and reveal the sender and website. 

We partially reconstruct a simulated Tor network, created using the algorithm described in Section~\ref{sec: Simulation}, using the packets generated in the simulation. Each jurisdiction (from a chosen set of collaborators) process their observed data to make partial circuits, which will be connected further to create full circuits. The jurisdictions output is verified by comparing it with the real circuits created in the simulation. We assume all Tor circuits has length five, this is not essential for our algorithm.

\subsection{Process observed packets}

Each jurisdiction process its observed packets to make partial circuits, using the classification discussed in Section~\ref{sec: Modeling jurisdictional adversaries}. All packets are labeled either as incoming, entering the jurisdiction, or outgoing, leaving the jurisdiction. We iterate over all incoming packets and try to combine each one with an outgoing packet. We only look at the outgoing packets which are close to the incoming packet, with respect to time. We first look for trivial connections: incoming or outgoing packets which can be classified as a \zero or~\emph{1} (packets passing though or packets ending inside the jurisdiction). If the incoming packet it not a trivial connection we look for an outgoing packet that share a common node with the incoming packet, that is, we try to make a \two connection. If there are no outgoing packets with a common node we look \three connections, where we want to find the outgoing packet which fits best with the incoming packet based on their timestamp difference. For \two and~\emph{3} we combine packets if their timestamp difference is close to some expected value, this value is the expected onion router delay and is derived from the family of lognormal distributions used in the simulation algorithm.

All connections made are stored as partial circuits, which contains the following information: a path of circuit nodes, four sets of timestamps, a probability score, and a list of circuit IDs. 

The four sets of timestamps shows the flow of data traveling over the path. Two of the sets represents packets traveling in one direction, say from left to right, where one contains incoming timestamps on the left side and one contains outgoing timestamps on the right side. Similar for the remaining two sets where the direction is opposite of the first two (from right to left). 

The probability score is used to evaluate how likely the circuit is part of a Tor circuit. The score is based on the time difference of the packets the partial circuit is made from. Partial circuits made from \zero and~\emph{1} connections have a score of zero. The score is the output of the probability density function of a lognormal distribution with the time difference as input. The distribution is derived from the family of lognormal distributions. The closer the difference is to the expected onion router delay the higher the score is. For \two and~\three connections we only connect an incoming packet with the outgoing packet which results in the highest score. A partial circuits probability score is equal to the sum of each of its packet pair's score.

To verify our data we use the Tor circuit IDs, where each circuit ID is collected from the packets used to make the partial circuit. Each Tor circuit has one unique ID, but a partial circuit can have more than one ID stored. Note that we do not use the circuit IDs during the reconstruction process, they are only used to verify the output.

When a jurisdiction has analyzed all of its packets we discard all partial circuits that are most like not part of a true Tor circuit, that is, if it has a low probability score. We set the discard limit based on trial and error, with a small discard limit we get a high false positive rate and with a large discard limit we get a low reconstruction rate.

\subsection{Process partial circuits}

Colluding jurisdictions share their partial circuits with each other to create full Tor circuits. We start by finding partial circuits that share two nodes, to find \four connections. Then we look for partial circuits that share one node, to make \five connections. Following by looking for partial circuits that have enough timestamps pairs, one from each partial circuit, that have a timestamp difference close to the expected value, that is, \six connections. If any new partial circuit was created while looking for these three types of connections we restart the partial circuit combination process, which is continued until there are no more circuits to combine.

We create a new partial circuit when we combine two, where we keep some of the data and discard some. We combine the two paths and any overlapping nodes are merged together. The new partial circuit only need four sets of timestamps, where we take two from the first partial circuit, say the two leftmost, and two from the second, say the two rightmost. We calculate a new probability score for the circuit, where we use the timestamps that is going to be discarded to calculate the score, using the same method we use to scoring packets. All circuit IDs contained in the two partial circuits are included in the new.

The reconstruction algorithm terminates when there are no more partial circuits to combine. The output is all partial circuits of length tree or longer, with a large enough probability score. Length five partial circuits are the potential full Tor circuits.

\subsection{Evaluate results}

From the simulation we store all Tor circuits created, including their circuit IDs, and we use this to check the output of the reconstruction algorithm. 

We compare the partial circuit's path with all simulated Tor circuits with a circuit ID that is equal to one of the partial circuit's stored IDs. The partial circuit is correct if the reconstructed path is equal to, or part of, one of the simulated path. That is, a partial circuit is considered correct if is its path is part of a simulated Tor circuit's path and it was created using packet data that was indeed sent over that Tor circuit. 

We split the output into two groups: incorrect partial circuits, showing the false positive rate, and correct partial circuits, from which we get the reconstruction rate and the relationship revealing rate. The reconstruction rate shows how much of the simulated Tor circuits the algorithm reconstructed, and the relationship revealing rate shows how many user-website connections we found.

\subsection{Runtime}

The main contributing factor to the runtime of the reconstruction algorithm is the number of packets, the second is the number of timestamps. The algorithm can be split into two parts: the first analyze packets to make partial circuits, the second analyze and combine partial circuits.

Let $\mathcal{J}$ denote the set of colluding jurisdictions, cooperating in analyzing packets and partial circuits. Each jurisdiction $J \in \mathcal{J}$ analyzes its incoming and outgoing packets to combine them and stores the timestamps. Let $p$ denote the number of incoming packets $J$ has stored, we assume $J$ also have $p$ outgoing packets. We look through all $p$ incoming packets and for each of them we compare it with some of the outgoing packets, we are only connect packets that have timestamps close to each other and keep a short list of outgoing packets with timestamps close to the current incoming packet's. Iterating over the packets is at most $O(p \log p)$. For each packet we connect we store its timestamp in a sorted list and inserting a timestamp is $O(p/k)$, where $k$ is the number of partial circuit made by jurisdiction $J$. The runtime of the packet analysis algorithm, for each jurisdiction $J$, is $O((p^2 \log p) / k)$, where $p$ is the number of packets (incoming or outgoing) $J$ has stored and $k$ is the number of partial circuits $J$ has made. The number of colluding jurisdictions $|\mathcal{J}|$ is a small constant. As an optimization, we store all partial circuits a jurisdiction makes and we only need to run the packet analysis algorithm for a jurisdiction once. (If we want to change the set of collaborating jurisdictions we do not have to redo the packet analysis every time.)

All partial circuits made by the colluding jurisdictions are analyzed in the second part of the reconstruction algorithm. Let $K$ denote the total number of partial circuits. When we combine partial circuits that share at least one common node we have a $O(K \log K)$ method for iterating through them, similar to the packet analysis method. However, when we combine partial circuits that do not share any common node we need a $O(K^2)$ method for iterating though them. We want to see if each partial circuit fits with all other, based on their combined path and timestamps. If two partial circuits' combined path is logical, looks like a Tor circuit, then we evaluate their timestamps and give a probability score for how well the two partial circuits fit together. Let $t$ denote the number of timestamps a partial circuit has. Evaluating the timestamps is $O(t \log t)$, using the method for packet analysis, and if the new probability score is high enough we create a new partial circuit and insert timestamps into sorted lists, which is $O(t)$. The runtime of the packet analysis algorithm is $O(K^2 t^2 \log t)$.

The wall-clock runtime of the reconstruction algorithm for: a simulated network with 1000 iterations and five jurisdictions is three hours, a simulated network with 10000 iterations and five jurisdictions is two and a half days.

\subsection{Improvements}

Our implementation is not perfect, and here we mention possible improvements to the reconstruction algorithm. The implementation is written in python, where an implementation written in a different language could be more efficient. Furthermore, each jurisdiction analyze their own data and can be run in parallel. 

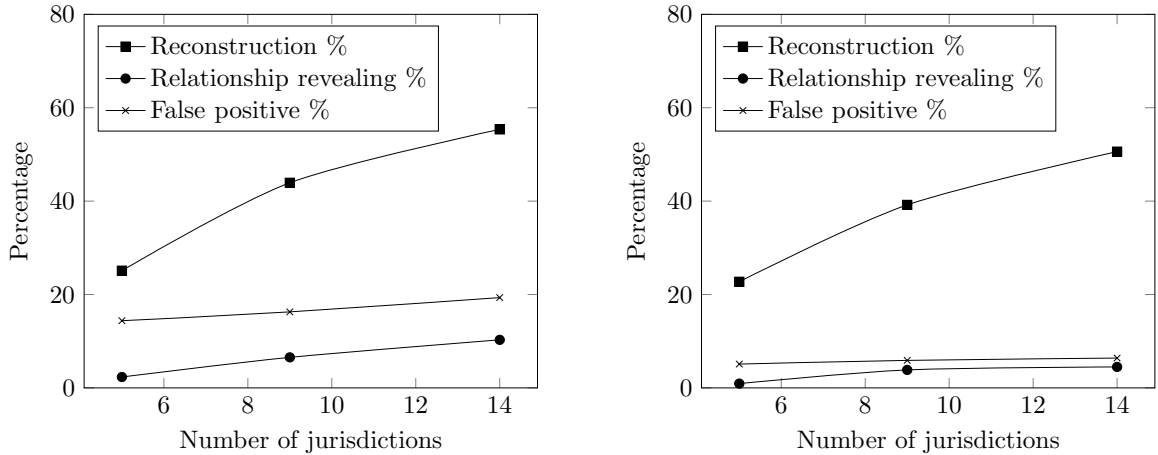
\begin{figure*}[tbp]
    \centering
    \begin{subfigure}[t]{0.49\textwidth}
	    \centering
		\begin{tikzpicture}[scale=0.87]
		\begin{axis}
		[
			xlabel={Number of jurisdictions},
			ylabel={Percentage},
			ymin=0,
			ymax=80,
			legend pos=north west,
			legend cell align={left}
		]
		\addplot[smooth, mark=square*] 
		coordinates {
			(5, 25.1)
			(9, 43.93)
			(14, 55.36)
		};
		\addplot[smooth, mark=*] 
		coordinates {
			(5, 2.33)
			(9, 6.54)
			(14, 10.29)
		};
		\addplot[smooth, mark=x] 
		coordinates {
			(5, 14.39)
			(9, 16.28)
			(14, 19.34)
		};
  		\legend{Reconstruction \%, Relationship revealing \%, False positive \%}
		\end{axis}
		\end{tikzpicture}
    \end{subfigure}%
    \hfill       
    \begin{subfigure}[t]{0.49\textwidth}
	    \centering
		\begin{tikzpicture}[scale=0.87]
		\begin{axis}
		[
			xlabel={Number of jurisdictions},
			ylabel={Percentage},
			ymin=0,
			ymax=80,
			legend pos=north west,
			legend cell align={left}
		]

		\addplot[smooth, mark=square*] 
		coordinates {
			(5, 22.73)
			(9, 39.19)
			(14, 50.60)
		};
		\addplot[smooth, mark=*] 
		coordinates {
			(5, 0.92)
			(9, 3.85)
			(14, 4.49)
		};
		\addplot[smooth, mark=x] 
		coordinates {
			(5, 5.10)
			(9, 5.89)
			(14, 6.39)
		};
  		\legend{Reconstruction \%, Relationship revealing \%, False positive \%}
		\end{axis}
		\end{tikzpicture}
    \end{subfigure}
    \begin{subfigure}[t]{0.48\textwidth}
	    \centering
       	\caption{Reconstructing a simulated network with a large number of iterations.}
       	\label{fig: Compare - big}
    \end{subfigure}%
    \hfill
    \begin{subfigure}[t]{0.48\textwidth}
	    \centering
       	\caption{Reconstructing a simulated network with a small number of iterations.}
       	\label{fig: Compare - small}
    \end{subfigure}
    \caption{Comparing reconstruction results. The more iterations used in the simulation the more data is generated, more data means more connections can be made -- both correct and false ones.}
    \label{fig: Compare}
\end{figure*}

\section{Reconstruction results}
\label{sec: Reconstruction results}

In the result we look at: the false positive rate, the reconstruction rate, and the relationship revealing rate. The false positive rate shows how many of the partial circuits are incorrect, the reconstruction and the relationship revealing rate only look at the partial circuit that are correct. The reconstruction rate shows how much of the simulated Tor circuits is reconstructed, partial circuits of length three or longer is used to find the reconstruction rate. The relationship revealing rate shows how many partial circuits that reveal the user-website connection.

\subsection{Results}
We simulate a Tor network, with a specified number of iterations, and use the output of the simulation algorithm as input to the reconstruction algorithm. We run the simulation algorithm once and the reconstruction algorithm several times. Every time we reconstruct we change the number of colluding jurisdictions. We include two reconstruction tests, in the first we look at how many simulation iterations affects the reconstruction results and in the second we look at how the coalition size affects the reconstruction results. The parameters used for the simulations are in Table~\ref{tab: Simulation parameters}, and the parameters used for the reconstructions are in Table~\ref{tab: Reconstruction parameters}.

In the first reconstruction test we compare reconstruction results from a simulation with a large number of iterations with reconstruction results from a simulation with a small number of iterations, see Figure~\ref{fig: Compare}. The average number of active Tor users in the simulation algorithm is close to the reported number of active users on Tor Metrics. The number of iterations specified for a simulation changes how many users have been active, but the average number of active should still be the same for all simulations. We see that all three rates are higher in the reconstruction results of the larger simulation. The larger simulation generates more data and the reconstruction algorithm has more to process, this means the reconstruction algorithm can make more connections and make more errors. By setting the cutoff bound for the probability score higher in the larger reconstruction we would get a lower false positive, reconstruction and relationship revealing rate, and get a result closer to the smaller reconstruction. Hence, having a larger number of iterations in the simulations only means longer computation time, and we will therefore only run the reconstruction algorithm on the smaller simulation for the second reconstruction test.

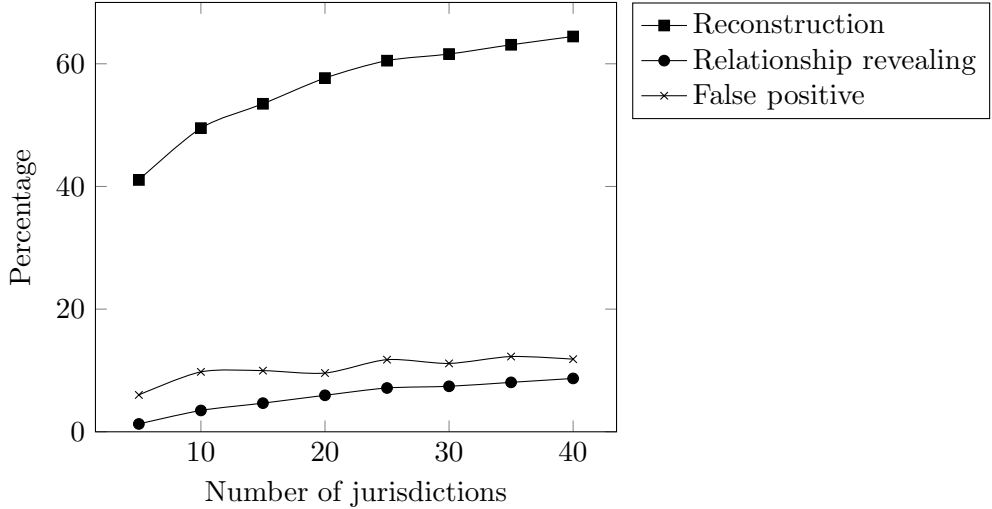
\begin{figure*}[tbp]
    \centering
	\begin{tikzpicture}
		\begin{axis}
		[
			xlabel={Number of jurisdictions},
			ylabel={Percentage},
			ymin=0,
			ymax=70,
			legend pos=outer north east,
			legend cell align={left}
		]
		\addplot[smooth, mark=square*] 
		coordinates {
			(5, 41.06)
			(10, 49.51)
			(15, 53.48)
			(20, 57.67)
			(25, 60.51)
			(30, 61.59)
			(35, 63.10)
			(40, 64.45)
		};
		\addplot[smooth, mark=*] 
		coordinates {
			(5, 1.28)
			(10, 3.48)
			(15, 4.67)
			(20, 5.95)
			(25, 7.14)
			(30, 7.42)
			(35, 8.06)
			(40, 8.70)
		};
		\addplot[smooth, mark=x] 
		coordinates {
			(5, 6.01)
			(10, 9.77)
			(15, 9.97)
			(20, 9.58)
			(25, 11.76)
			(30, 11.15)
			(35, 12.27)
			(40, 11.84)
		};
  		\legend{Reconstruction, Relationship revealing, False positive}
		\end{axis}
	\end{tikzpicture}
    \caption{Reconstruction results for an increasing number of jurisdictions.}
    \label{fig: Coalition size}
\end{figure*}

In the second reconstruction test we look at how the coalition size affects the reconstruction results, see Figure~\ref{fig: Coalition size}. The simulated network is always the same, we only changes the number of jurisdictions cooperating for each run of the reconstruction algorithm. After the simulation algorithm is completed the jurisdictions are sorted based on the amount of data they have stored, from big to small. The jurisdiction coalition chosen for the $n$'th run of the reconstruction algorithm is the first $5n$ jurisdictions in the sorted list. This means that for each run of the reconstruction algorithm the coalition consists of the jurisdictions used in the previous run plus five new ones. For each new run of the reconstruction algorithm the five new jurisdictions has less and less data to contribute to the coalition, hence, all three rates flattens out and stabilizes.

We can only speculate as to why the reconstruction rate peak at 65 percent and the relationship rate peaks at 10 percent for the second reconstruction test. This is partly of how the Tor network builds circuits and partly because of our implementation. If a Tor circuit's traffic doesn't cross the border of a jurisdiction, then no data is recorded and can never be reconstructed. The jurisdictions used for the reconstruction simply does not observe enough data. Furthermore, we believe that each jurisdiction discards too much information when analyzing their own data. When we combine incoming and outgoing traffic we get a pile of leftovers, traffic that has not been used to reconstruct. A better algorithm could possibly reduce the amount of leftover traffic and find ways to use the leftovers.

\section{Discussions}
\label{sec: Disscussions}

\subsection{Path selection countermeasure} 
\label{sec: Path selection countermeasure}

To be able to reconstruct a Tor circuit the jurisdictional adversaries needs to observe traffic sent to and from the user and the website. If the traffic sent from the user to the guard node is does not cross a jurisdictional border then no traffic can be observed and the circuit can never be fully reconstructed, similar for when the exit node and the website communicates. If the user specify its Tor circuit such that the traffic that crosses the jurisdictional borders is sent between onion routers then the adversaries cannot see the user or the website and are never able to connect them. 

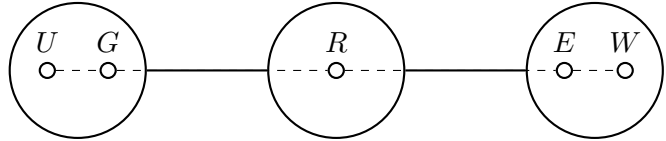
\begin{figure}[tbp]
	\centering
	\begin{tikzpicture}
		\tikzstyle{leaf}=[circle,draw,inner sep=2pt,thick]
		\tikzstyle{jur}=[circle,draw,inner sep=18pt,thick]

		\node[leaf, label = {90:{$U$}}] (U) at (-3.8,0) {};
		\node[leaf, label = {90:{$G$}}] (G) at (-3,0) {};
		\node[leaf, label = {90:{$R$}}] (R) at (0,0) {};
		\node[leaf, label = {90:{$E$}}] (E) at (3,0) {};
		\node[leaf, label = {90:{$W$}}] (S) at (3.8,0) {};
		
		\node[jur] (j1) at (-3.4,0) {};
		\node[jur] (j2) at (0,0) {};
		\node[jur] (j3) at (3.4,0) {};
		
		\draw[dashed] (U) -- (G) -- (R) -- (E) -- (S); 
		\draw[thick] (j1) -- (j2) -- (j3); 
	\end{tikzpicture}
	\caption{Path selection where the jurisdictional adversaries are unable to connect the user $U$ with the website $W$, since they can only observe traffic sent between the guard node $G$, the relay node $R$, and the exit node $E$.}
	\label{fig: Avoid}
\end{figure}

The following Tor circuit selection prevents the jurisdictional adversaries breaking the relationship anonymity. A user $U$ wants to connect to a website $W$. The user chooses the onion routers as follows: the guard node $G$ should be situated in the same jurisdiction as the user $U$, the relay node $R$ can be in any jurisdiction, and the exit node $E$ should be located inside the same jurisdiction as the website $W$. See Figure~\ref{fig: Avoid}.

Note that this path selections only avoids the jurisdictional adversaries, it is possible that other types adversaries could break the relationship anonymity if the users use this path selections. For example, an adversary corrupting single onion routers can see traffic sent inside a jurisdiction if a circuit visits a corrupted node, and can possibly see the user or website of the circuit.

\subsection{Passive global adversaries} 
\label{sec: Passive global adversaries}

We claim that the best attack the jurisdictional adversary can do is to passively observe Tor traffic, and a coalition of jurisdictions are therefore a passive global adversary versus the Tor network. 

Tor uses a TLS connection between circuit nodes (except between the exit node and the website), which provides confidentiality and message integrity~\cite{RFC5246} and implies that Tor is IND--CCA~\cite{KY01}. Any active attack against messages sent between circuit nodes will be detected and prevented. The best attack the adversaries can do is to passively observe Tor traffic, and possibly stop traffic.

The jurisdictional adversaries indirectly monitor all onion routers inside its jurisdiction. If the jurisdictional adversaries cooperates in reconstructing Tor circuits they quickly become global, since they monitor a large portion of the onion router. In addition, a large set of jurisdictions has the power to reveal the relationship of a circuit if they choose to do so. 


\bibliographystyle{plain}
\bibliography{torbib}

\newpage

\appendix

\section{Simulation and reconstruction parameters}
\label{app: Simulation and reconstruction parameters}

\begin{table}[h!]
\centering
\caption{Parameters and data of the simulations used in the results.}
\label{tab: Simulation parameters}
\begin{tabular}{llll}
\toprule
 & Iterations & Packets sent & Data stored \\
\midrule
Small simulation & $1000$ & $14,001,560$ & $0.69$ GB \\
Large simulation & $10000$ & $141,404,190$ & $5.16$ GB \\
\bottomrule
\end{tabular}
\end{table}

\begin{table}[h!]
\centering
\caption{Parameters and data of the reconstruction algorithm.}
\label{tab: Reconstruction parameters}
\begin{tabular}{llcc}
\toprule
 &  & \multicolumn{2}{c}{Probability score lower bound} \\
 & Jurisdictions & \two & \three \\
 \midrule
Figure~\ref{fig: Compare} & \begin{tabular}[c]{@{}l@{}}US, GB, AU, CA, NZ,\\ DK, FR, NL, NO,\\ BE, DE, IT, ES, SE\end{tabular} & 10 & 20 \\ \midrule
Figure~\ref{fig: Coalition size} & \begin{tabular}[c]{@{}l@{}}DE, US, FR, RU, GB,\\ PL, NL, UA, JP, DK,\\ CN, CA, NO, BG, AE,\\ IT, SE, CH, TR, GR,\\ FI, AT, RO, MD, CZ,\\ ID, ES, PT, IN, IS,\\ HU, BE, IE, BR, AU, \\ SG, SC, TH, SK, LU,\\ HR, HK, PK, MY, LV,\\ DJ, ZA, KR, IR, EG,\\ LT, VE, VN, MX, TW,\\ PH, AR, CR, IL, EE,\\ SA, PA, CL, SI, CO,\\ BN, KZ, GE, BY, CY,\\ AZ, MA, AO, OM, NZ, \\ RS, CW, KG, CM, JO,\\ KE, BD, ZM, KN, SN,\\ QA, NA, TZ, BA, DO, \\ UZ, AL, AM, MN, LK,\\ GH, SD, MZ, PY, KW\end{tabular} & 10 & 20 \\
\bottomrule
\end{tabular}
\end{table}

\end{document}